\documentclass[journal=jctcce,manuscript=article]{achemso}
\usepackage{color}
\usepackage{array}
\usepackage{mathtools}
\usepackage{multirow}
\usepackage{amsmath}
\usepackage{amssymb}
\usepackage{graphicx}

\makeatletter

\usepackage{chemformula}
\usepackage[version=3]{mhchem}
\usepackage{xcolor}
\usepackage{longtable}
\usepackage{tablefootnote}
\usepackage{tcolorbox}
\usepackage{enumerate}
\usepackage{multirow}
\usepackage{subcaption}

\author{Chaoqun Zhang}
\altaffiliation{These authors contributed equally to this work}
\author{Christian Venturella}
\altaffiliation{These authors contributed equally to this work}
\author{Enzhi Chen}
\author{Tianyu Zhu}
\email{tianyu.zhu@yale.edu}
\affiliation{Department of Chemistry, Yale University, New Haven, CT 06520, USA}

\title
  {Transferable Machine Learning of Electronic Hamiltonians with Superposition-of-Atomic-Potentials Features}

\makeatother

\begin{document}

\begin{abstract}
Machine learning (ML) of electronic Hamiltonians offers a unified route to electronic wave functions and physical observables.
We introduce a Hamiltonian learning framework built on electronic features derived from the superposition-of-atomic-potentials (SAP) approximation, an efficient self-consistent-field initial guess that captures essential electron-electron screening. 
SAP quantities define a symmetry-adapted intrinsic atomic orbital learning basis and 
provide physics-informed inputs to an orbital-based graph neural network that predicts converged Kohn-Sham Fock matrices.
To extend the approach to larger basis sets, we further develop a downfolding scheme that predicts large-basis electronic structure from minimal-basis features. 
On the QM9 dataset, the model accurately reproduces frontier and core orbital energies, dipole moments, and the full density of states. 
For organic charge-transport materials, it yields accurate intermolecular transfer integrals for benzene, tetracyanoquinodimethane (TCNQ), and tetrathiafulvalene (TTF) dimers, and transfers to unseen substituted-benzene heterodimers with a mean absolute error of 4.8 meV.
These results establish SAP-based ML of electronic Hamiltonians as a transferable and scalable tool for high-throughput electronic-structure prediction.

\end{abstract}

\section{Introduction}
Data-driven machine learning (ML) has emerged as a powerful tool for modeling molecules and materials \cite{butlerMachineLearningMolecular2018,vonlilienfeldExploringChemicalCompound2020,keithCombiningMachineLearning2021,deringerGaussianProcessRegression2021,tangDeepEquivariantNeural2024,linDeeplearningAtomisticSemiempirical2025,kingCartesianEquivariantRepresentations2025}. 
By learning from reference quantum chemistry data, ML models can reproduce the results of expensive first-principles calculations at a small fraction of the cost, bridging the gap between accuracy and computational efficiency. 
This capability has driven broad adoption across chemistry and materials science. 
ML interatomic potentials now reach near-ab-initio accuracy on potential energy surfaces and enable molecular dynamics simulations at length and time scales far beyond the reach of direct ab initio methods \cite{behlerGeneralizedNeuralNetworkRepresentation2007,zhangDeepPotentialMolecular2018,batatiaMACEHigherOrder2022,batznerE3equivariantGraphNeural2022,batatiaFoundationModelAtomistic2025,woodUMAFamilyUniversal2026,liPredictiveFreeEnergy2026}. 
In parallel, ML models have been used to predict a wide range of molecular and materials properties, including charges, dipole moments and polarizabilities \cite{unkePhysNetNeuralNetwork2019,qiaoInformingGeometricDeep2022,sunMolecularDipoleMoment2022,kingMachineLearningCharges2025} as well as spectroscopic responses \cite{westermayrPhysicallyInspiredDeep2021,cignoniElectronicExcitedStates2024,houMBFormerGeneralTransformerbased2025}.
Coupled with generative and inverse-design strategies \cite{sanchez-lengelingInverseMolecularDesign2018}, these models have accelerated high-throughput screening and the rational design and discovery of next-generation functional materials.

Most existing ML models in quantum chemistry are trained to reproduce individual observables, such as energies or electron densities.
Predicting multiple properties beyond the specific training target remains challenging, typically requiring a separate model for each new quantity.
Such property-specific learning limits transferability across tasks and do not fully exploit the underlying operator structure of electronic structure theory. 
In contrast, targeting more fundamental quantum chemical quantities, such as the semi-empirical Hamiltonian parameters \cite{dralMachineLearningParameters2015}, electronic Hamiltonian \cite{schuttUnifyingMachineLearning2019}, electronic densities \cite{grisafiTransferableMachineLearningModel2019,shaoMachineLearningElectronic2023}, and electronic wave functions \cite{houUnsupervisedRepresentationLearning2024,rathInterpolatingNumericallyExact2025}, provides direct access to a broad range of derived electronic properties.
Motivated by these advantages, recent years have seen rapid developments in machine-learned Hamiltonians \cite{liDeeplearningDensityFunctional2022,nigamEquivariantRepresentationsMolecular2022,zhouDeepLearningDynamically2022,zhongTransferableEquivariantGraph2023,guDeepLearningTightbinding2024,venturellaMachineLearningManyBody2024,yuQH9QuantumHamiltonian2024,sumanExploringDesignSpace2025,venturellaUnifiedDeepLearning2025}, covering mean-field Fock and many-body quasiparticle Hamiltonians for both molecular and periodic systems.
The successful extension of these models to downstream tasks, such as the prediction of electron-phonon couplings \cite{liDeepLearningDensityFunctional2024,zhongAcceleratingCalculationElectron2024}, further demonstrates the robustness and broad capabilities of Hamiltonian-targeted learning.

A central, remaining challenge in learning Hamiltonians lies in designing input features that are simultaneously transferable and computationally efficient. 
While purely geometric descriptors are inexpensive to compute, their predictive performance often degrades during chemical and conformational extrapolation. 
Consequently, recent efforts have explored electronically informed features, such as external nuclear potentials ($V_{\text{nuc}}$) \cite{nigamMachineLearningElectronic2026} or one-shot Fock matrices \cite{shakibaMachineLearnedKohnSham2024}.
However, bare external potentials neglect two-electron screening effects and become numerically unbounded as system size increases, while a full one-step Fock build introduces significant computational overhead.
As a balanced alternative, we propose utilizing interaction-matrix features derived from the superposition-of-atomic-potentials (SAP) approximation \cite{lehtolaAssessmentInitialGuesses2019,lehtolaEfficientImplementationSuperposition2020}.
Originally designed as a low-cost initial guess for density functional theory (DFT), SAP provides a robust, electronically informed representation that inherently captures electron-electron screening.
By leveraging the SAP operator both to construct a symmetry-adapted learning basis and to generate input features, we develop an accurate and transferable Hamiltonian-learning model.

A second major challenge for matrix-to-matrix learning is scalability. 
Extending predictions to larger systems and basis sets rapidly increases both data storage requirements and model complexity. 
To address this bottleneck, we introduce a downfolding framework that enables the prediction of large-basis electronic structures using features with minimal-basis dimensions.
Together, the SAP-derived features and the downfolding strategy establish a practical pathway toward transferable, scalable ML Hamiltonian models for quantum chemistry. 
We first demonstrate the efficacy of this framework on the QM9 dataset \cite{ramakrishnanQuantumChemistryStructures2014}, accurately reproducing frontier-orbital energies, dipole moments, carbon 1s core-level shifts, and the density of states. We further apply the framework to intermolecular charge-transfer integrals in organic semiconductors by training and testing a separate model on dimers of benzene, tetracyanoquinodimethane (TCNQ), and tetrathiafulvalene (TTF), which generalizes accurately to substituted-benzene heterodimers that are excluded from the training set.

\section{Theory and Computational Details}
\label{sec:theory}
\subsection{DFT features with superposition of atomic potentials}
The superposition of atomic potentials (SAP) approach \cite{lehtolaAssessmentInitialGuesses2019,lehtolaEfficientImplementationSuperposition2020} has recently been shown to provide an efficient and robust initial guess for mean-field calculations.
The SAP Hamiltonian is an effective one-electron operator with a screened electron–nucleus interaction,
\begin{equation}
    H^{\text{SAP}} = \sum_i\left[-\frac{\nabla_i^2}{2} - \sum_A \frac{Z_\text{{eff}}(r_{iA})}{r_{iA}} \right]
\end{equation}
where $r_{iA}$ is the distance between electron $i$ and nucleus $A$.
In practice, the distance-dependent nuclear charge $Z_\text{eff}(r_{iA})$ is fitted to pre-tabulated spherical atomic densities, enabling efficient evaluation of matrix elements.

In this work, training features are constructed from DFT quantities obtained via one-shot diagonalization of $H^{\text{SAP}}$.
We include the exchange correction from the local density approximation (LDA) in the SAP Hamiltonian \cite{lehtolaAssessmentInitialGuesses2019}, as implemented in PySCF \cite{sunPySCFPythonbasedSimulations2018,sunPythonSimulationsChemistry2026}.
The SAP method is significantly less expensive than alternative initial guesses, such as the superposition of atomic densities (SAD) or projection from minimal basis sets, as it directly yields approximate Fock matrices without requiring tensor contractions over molecular two-electron integrals.
\begin{figure}[h]
    \centering
\includegraphics[width=1\linewidth]{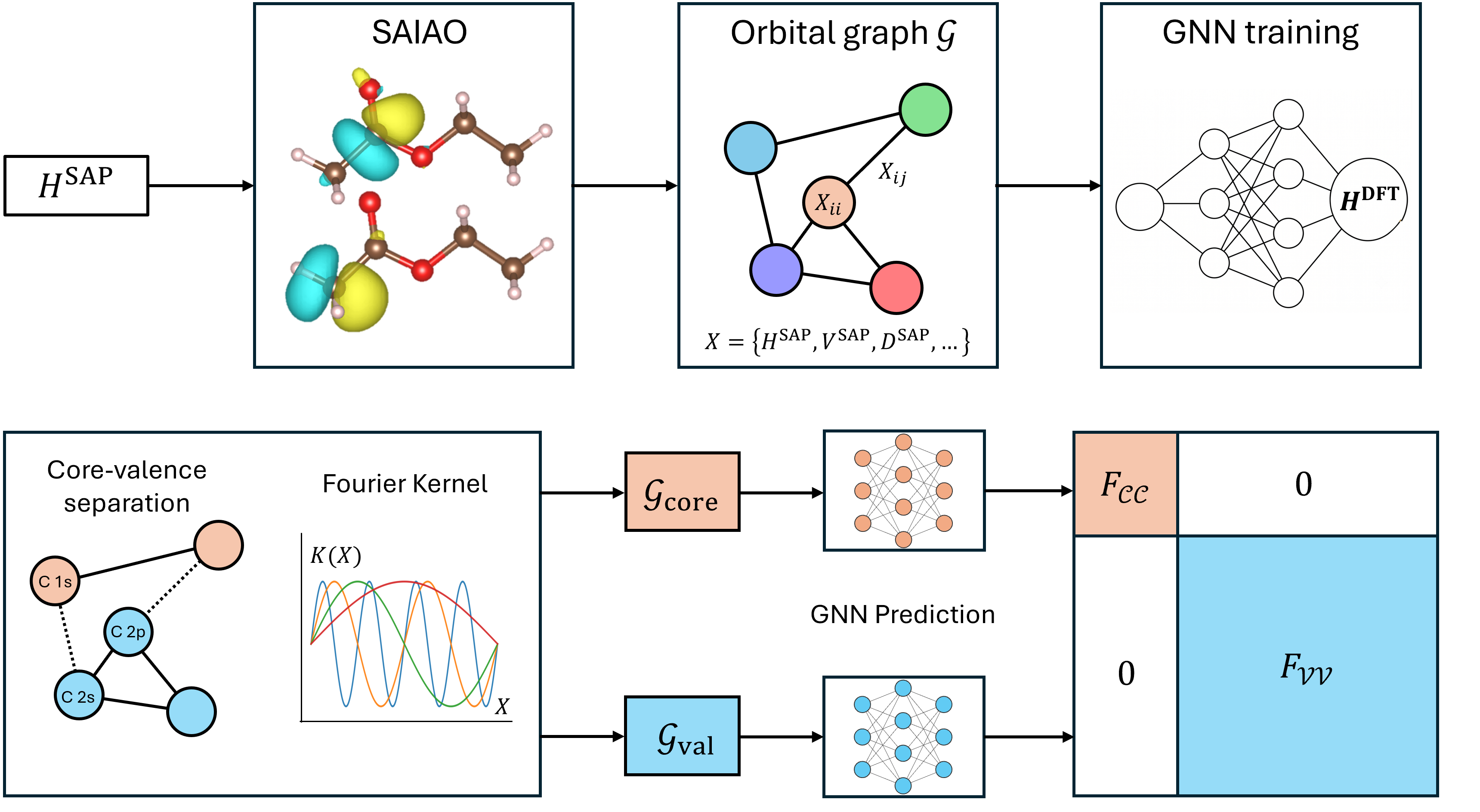}
\caption{Machine learning framework for predicting DFT Hamiltonian. 
Top panel: Equivariant SAP features are constructed in the SAIAO basis and mapped onto the orbital graph as ML input. The converged DFT Fock matrices are predicted from the graph neural network (GNN).
Bottom panel: Orbital graphs are decomposed into core and valence blocks. Each block is assigned its own GNN and trained jointly. A Fourier feature encoder is implemented to complement standard SAP features.}
\label{fig:ml scheme}
\end{figure}

$H^{\text{SAP}}$ can be regarded as an approximate Fock matrix and is used to generate a local basis set for machine learning.
As shown in Fig. \ref{fig:ml scheme} (top panel), and following our previous work on ML many-body Green's functions~\cite{venturellaMachineLearningManyBody2024,venturellaUnifiedDeepLearning2025},
we construct intrinsic atomic orbital plus projected atomic orbital basis (IAO+PAO)~\cite{kniziaIntrinsicAtomicOrbitals2013,cuiEfficientImplementationInitio2020} using occupied molecular orbitals (MO) obtained from $H^{\text{SAP}}$.
This is followed by symmetry adaptation by diagonalizing atomic angular momentum blocks \cite{qiaoOrbNetDeepLearning2020}.
Both the SAP features and the ML targets (converged DFT Fock matrices) are rotated into the resulting symmetry-adapted intrinsic atomic orbital plus projected atomic orbital basis, which we refer to as the SAIAO basis.
An orbital graph is then constructed by mapping diagonal and off-diagonal matrix elements to nodes and edges, respectively.

\subsection{Machine learning architecture}
Our model builds on the graph neural network (GNN) introduced in our
previous work for learning many-body Green's functions \cite{venturellaMachineLearningManyBody2024,venturellaUnifiedDeepLearning2025}. 
In contrast to those works, the learning target in this work is the static
Kohn-Sham Fock matrix.
The architecture therefore carries no frequency-dependent features.
The present work features two methodological advances (Fig. \ref{fig:ml scheme} bottom): (i) a multi-block decomposition that learns
the core and valence sectors of the Fock matrix with separate sub-networks,
and (ii) a sinusoidal (Fourier) feature encoder to complement the standard DFT features. 

\paragraph{Multi-block core/valence learning.}
The central architectural element is a block decomposition of the Fock
matrix according to a core-valence separation in SAIAO basis.
Each local orbital is labeled as core or
valence, partitioning the orbital index set into
$\mathcal{C}$ and $\mathcal{V}$. The Fock matrix is correspondingly
partitioned into core-core, valence-valence, and core-valence blocks,
\begin{equation}
  F^{\text{SAIAO}}  =
  \begin{pmatrix}
    F_{\mathcal{C}\mathcal{C}}^{\text{SAIAO}}  & F_{\mathcal{C}\mathcal{V}}^{\text{SAIAO}} \\[2pt]
    F_{\mathcal{V}\mathcal{C}}^{\text{SAIAO}}  & F_{\mathcal{V}\mathcal{V}}^{\text{SAIAO}} 
  \end{pmatrix}.
\end{equation}
Because the localized core orbitals couple weakly to the valence space, the off-diagonal core-valence block is neglected ($F_{\mathcal{C}\mathcal{V}}\approx 0$), and only the diagonal blocks are predicted.

Each block is assigned its own GNN with independent hyperparameters, allowing the network capacity to be tailored to the complexity of each sector.
In this work the valence block uses larger 
embeddings and more message-passing rounds than the comparatively rigid
core block (e.g., $128$ vs.\ $64$ message-passing channels and $4$ vs.\ $2$
update rounds). 
The two blocks are trained jointly with a per-block loss, ensuring that the large-magnitude but chemically inert core elements do not dominate optimization.
For the systems studied in this work, the core orbitals include the 1s orbitals of C, N, O, F, and the 1s, 2s, and 2p orbitals of S, while all remaining occupied and virtual orbitals are assigned as valence space.

\paragraph{Fourier feature encoder.}
To resolve the multiple length and energy scales present in the input, each scalar input feature $x$ is expanded in a
sinusoidal embedding before entering the encoder,
\begin{equation}
  \gamma(x) = \Big[\, x,\;
    \{\sin(x/s_k)\}_{k=1}^{K},\;
    \{\cos(x/s_k)\}_{k=1}^{K}\,\Big],
\end{equation}
with fixed scales $\{s_k\}$ (here $K=4$, $s_k\in\{1,0.5,0.1,0.05\}$),
mapping a $d$-dimensional feature vector to dimension $d(2K+1)$. 

\paragraph{Encoder, message passing, and decoder.}
Node and edge features are first transformed by a residual multilayer perceptron with SiLU activations (the encoder), and the resulting embeddings are refined by $T$ rounds of attentional message passing \cite{venturellaUnifiedDeepLearning2025}, in which learned attention scores weight the messages aggregated at each node and a residual update is applied to both node and edge embeddings. 
Two separate decoders map the final node and
edge embeddings to the diagonal and off-diagonal
Fock-matrix elements, respectively, which are assembled into a symmetric matrix.

\paragraph{Loss function.}
The network is trained with a composite, physics-informed loss evaluated
per molecule (and, for the multi-block model, per orbital block). The
primary term is the mean-squared error (MSE) between the predicted ($\hat{F}^{\text{SAIAO}}$) and reference ($F^{\text{SAIAO}}$) Fock matrix elements in the SAIAO basis,
\begin{equation}
  \mathcal{L}_{F} = \mathcal{L}_{MSE}(\hat{F}^{\text{SAIAO}}, F^{\text{SAIAO}}),
\end{equation}
To enforce physically consistent predictions, additional penalties are included,
\begin{equation}
  \mathcal{L} = \mathcal{L}_{F}
  + \beta_{\epsilon} \mathcal{L}_{MSE}(\hat{\epsilon}, \epsilon)
  + \beta_{P}\mathcal{L}_{MSE}(\hat{P}^{\text{SAIAO}}, P^{\text{SAIAO}})
  + \beta_{\mu}\mathcal{L}_{MSE}(\hat{\mu}, \mu),
\end{equation}
where the auxiliary terms penalize errors in molecular orbital energies ($\epsilon$), density matrices in SAIAO basis ($P^{\text{SAIAO}}$), and molecular dipole moments ($\mu$), respectively. Here, the predicted quantities $\hat{\epsilon}$, $\hat{P}^{\text{SAIAO}}$, and $\hat{\mu}$ are obtained directly from the ML-predicted Fock matrix.
$\beta_{\epsilon},\beta_{P},\beta_{\mu}$ control the relative weights of the extra penalty terms.
Unless otherwise stated, we set $\beta_{\epsilon}=0.1$ and $\beta_{P}=\beta_{\mu}=0.05$ for the valence block, while only $\mathcal{L}_{F}$ term is applied to the core block.

\subsection{Hamiltonian downfolding}
High-lying virtual orbitals contribute weakly to ground-state properties, motivating downfolding approaches that reduce the Hamiltonian dimension while preserving accuracy.
Learning the downfolded Hamiltonian with accordingly downfolded input features therefore provides a promising route to extend Hamiltonian learning to larger systems and basis sets~\cite{nigamMachineLearningElectronic2026}.

The Hamiltonian downfolding scheme used in this work is inspired by L\"owdin partitioning \cite{lowdinStudiesPerturbationTheory1962} and exact two-component (X2C) theories \cite{dyallInterfacingRelativisticNonrelativistic2001,kutzelniggQuasirelativisticTheoryEquivalent2005,iliasInfiniteorderTwocomponentRelativistic2007}.
Writing the Hamiltonian in block form for subsystems $A$ and $B$,
\begin{equation}
    \left[ \begin{array}{cc}
        H^{AA} & H^{AB} \\
        H^{BA} & H^{BB}
    \end{array} \right]
    \left[ \begin{array}{c}
        C^{A}  \\
        C^{B} 
    \end{array} \right]=E
    \left[ \begin{array}{c}
        C^{A}  \\
        C^{B} 
    \end{array} \right]
\end{equation}
one obtains an energy-dependent relation $C^B=(E-H^{BB})^{-1}H^{BA}C^A$.
Defining a matrix $X$ such that $C^B=XC^A$,
the eigenvalue equation can be reduced to an $A$-only form. 
After normalization,
\begin{equation}
    H^{\text{eff}}C^{\text{eff}}=EC^{\text{eff}}
\end{equation}
with
\begin{equation}
\label{eqn:Heff}
    H^{\text{eff}} = R^\dagger\left[H^{AA}+H^{AB}X+X^\dagger H^{BA} + X^\dagger H^{BB}X\right] R
\end{equation}
\begin{equation}
    C^{\text{eff}} = R^{-1}C^A
\end{equation}
\begin{equation}
    R = \left[I + X^\dagger X\right]^{-1/2}    
\end{equation}

The $X$ matrix is obtained numerically from the first $N_A$ eigenvectors as $X=C^B C^{A\dagger}(C^AC^{A\dagger})^{-1}$.
When constructed from the exact DFT Hamiltonian, $X^{\text{DFT}}$ reproduces the corresponding eigenvalues, orbitals, and derived properties.
We mention that one can also formulate the same downfolding in a non-orthogonal basis set using a proper normalization matrix $R$. We refer the readers to Ref.\citenum{liuExactTwocomponentHamiltonians2009} for details. 

In the ML framework, we distinguish between \emph{downfolding learning} (with downfolding) and \emph{direct learning} (without downfolding).
In downfolding learning, SAIAOs are constructed from $H^{\text{SAP}}$ in the larger basis, and $X^{\text{SAP}}$ is computed from its eigenvectors.
Features are transformed to a minimal basis using $X^{\text{SAP}}$, while the DFT Hamiltonian target is downfolded using $X^{\text{DFT}}$ to preserve the orbital spectrum.
This introduces a basis-set mismatch between features and targets.
After feature and target construction, orbital graphs are built in the minimal basis and training proceeds as in direct learning.


\subsection{Calculations of charge transfer integrals}
Charge transport in organic semiconductor materials can be described using a simple tight-binding model with an electronic part
\begin{equation}
    H^{el} = \sum_I \varepsilon_I a_I^\dagger a_I + \sum_{I\neq J } t_{IJ} a_I^\dagger a_J,
\end{equation}
where $\varepsilon_I$ and $t_{IJ}$ denote site energies and intermolecular transfer integrals.
These parameters describe the polarization and the electronic coupling between adjacent molecules.
The basis functions $I,J,\dots$ correspond to charge-localized diabatic states.
Within a mean-field approximation, these quantities may be expressed in terms of one-electron orbital interactions.
One commonly used approach \cite{valeevEffectElectronicPolarization2006} evaluates the interaction Hamiltonian from DFT calculations on the dimer ($D$) and the isolated monomers ($I$ and $J$),
\begin{equation}
\label{eqn:tilde H}
    \mathbf{\tilde{H}} = \left[ \begin{array}{cc}
        \tilde{e}_I & \tilde{t}_{IJ} \\
        \tilde{t}_{JI} & \tilde{e}_J
    \end{array} \right]
\end{equation}
The matrix elements are given by
\begin{equation}
    \tilde{e}_I = \langle \Psi_I | H^D | \Psi_I \rangle = \sum_{\mu\nu \in I} C_{\mu}^{I\ast} H_{\mu\nu}^D C_{\nu}^I
\end{equation}
\begin{equation}
    \tilde{t}_{IJ} = \langle \Psi_I | H^D | \Psi_J \rangle = \sum_{\mu\in I,\nu \in J} C_{\mu}^{I\ast} H_{\mu\nu}^D C_{\nu}^J
\end{equation}
where $H^D_{\mu\nu}$ is the Fock matrix element in the AO basis for the dimer system, $C_\mu^I$ and $C_\mu^J$ are the MO coefficients for isolated monomer $I$ and $J$, respectively.
In this work, HOMO (LUMO) orbitals are considered for electron-rich (electron-poor) systems. The tight-binding Hamiltonian is obtained via L\"owdin orthonormalization,
\begin{equation}
    \mathbf{H} =\left[ \begin{array}{cc}
        e_I & t_{IJ} \\
        t_{JI} & e_J
    \end{array} \right]= \mathbf{\tilde{S}}^{-1/2} \mathbf{\tilde{H}} \mathbf{\tilde{S}}^{-1/2}
\end{equation}
where $\mathbf{\tilde{S}}$ is the orbital overlap matrix with $\tilde{S}_{II}=1$ and $\tilde{S}_{IJ} = \sum_{\mu\in I,\nu \in J} C_{\mu}^{I\ast} S_{\mu\nu}^D C_{\nu}^J$.

For ML predictions, the model provides DFT Hamiltonians for both dimers and monomers.
Monomer wavefunctions are obtained by diagonalizing the predicted monomer Hamiltonians.
Accurate transfer integrals therefore require balanced accuracy across dimers and monomers, as well as intra- and intermolecular interactions.

\section{Results and Discussion}
\label{sec:results}
\subsection{QM9 Benchmark}

For the benchmark tests, we performed DFT calculations on all molecules in QM9 dataset \cite{ramakrishnanQuantumChemistryStructures2014} using PBE0 functional \cite{adamoReliableDensityFunctional1999}. 
The cc-pVDZ basis set \cite{ccpvxzBtoNe} was used to generate reference data for direct learning models, whereas the cc-pVTZ basis set was employed for downfolding learning models. 
Training sets were randomly sampled from the full QM9 database.

For direct learning models, the SAP Fock matrix, kinetic energy, density matrices derived from the SAP Fock, and orbital Boys locality were used as node features. 
The orbital Boys locality for orbital $ \phi_i$ is defined as $f_i\equiv\langle \phi_i|\vec{r}^2|\phi_i\rangle - |\langle \phi_i|\vec{r}|\phi_i\rangle|^2 $ where $\vec{r}$ is the electronic position operator.
The SAP Fock, SAP effective two-electron potential ($H^{\text{SAP}}-h^{\text{core}}$), kinetic energy, and density matrices were used as edge features.
Combined with the Fourier components, a total of 36 node features and 36 edge features were used.
A $\Delta$-learning strategy was used for direct learning, in which the models were trained to predict the difference between SAP Fock matrix and the reference PBE0 Fock matrix.

For downfolding learning models, we added $H^{AA}$ and $H^{AB}X+X^\dagger H^{BA} + X^\dagger H^{BB}X$ in Eq.~\ref{eqn:Heff} to both node and edge features to complement other downfolded properties.
Due to the mismatch of the downfolding $X$ matrices, the $\Delta$-learning strategy is ill-defined for this approach. The downfolded PBE0 Fock matrices were thus used as ML targets.

\subsubsection{Frontier orbital energies and dipole moments}
\begin{figure}[h]
    \centering
\includegraphics[width=1\linewidth]{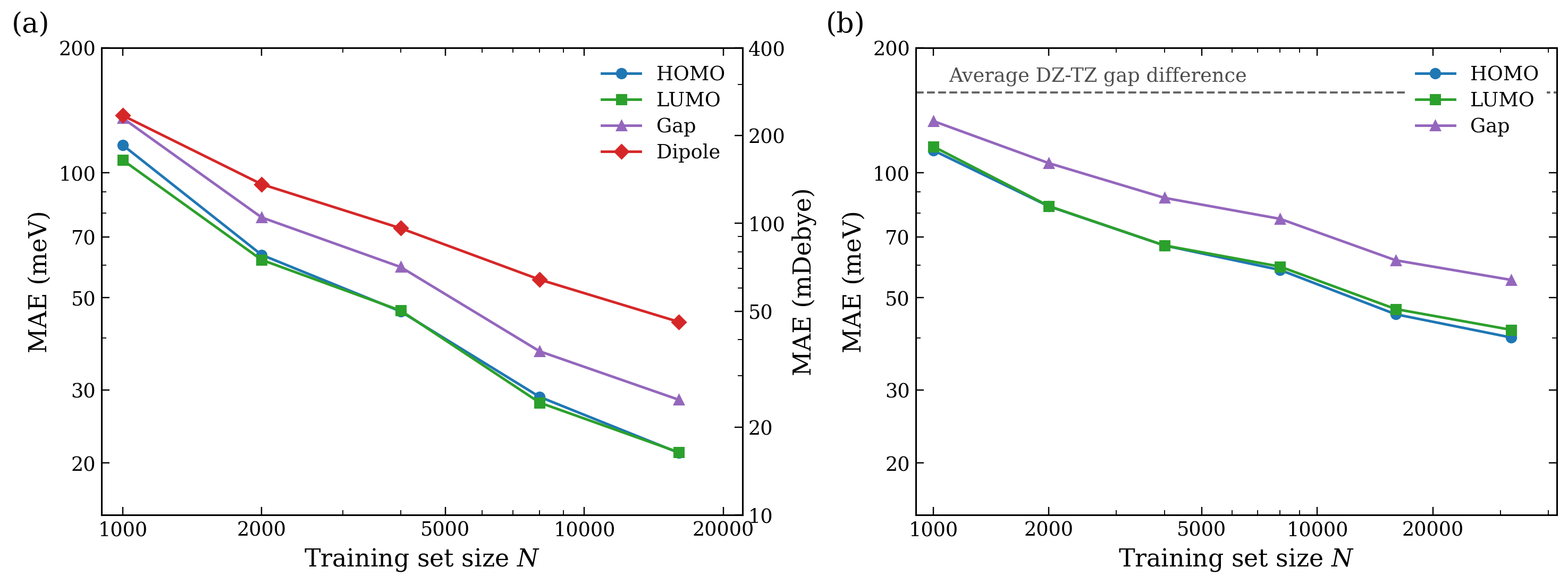}
\caption{Learning curves on the QM9 dataset as a function of training set size $N$.
(a) Direct learning of cc-pVDZ reference quantities: MAEs for the HOMO energy, LUMO energy, and HOMO-LUMO gap are shown on the left axis (meV), while the MAE for the molecular dipole moment is shown on the right axis (mDebye).
(b) Downfolding learning of cc-pVTZ reference quantities: MAEs for the HOMO energy, LUMO energy, and HOMO-LUMO gap.
The dashed horizontal line marks the mean HOMO-LUMO gap difference between true cc-pVDZ and cc-pVTZ DFT references across QM9 (157 meV).}
\label{fig:qm9}
\end{figure}

We benchmark the data efficiency of the proposed ML approach on the HOMO and LUMO orbital energies, HOMO-LUMO energy gap, and molecular dipole moment, by training on subsets of increasing size and evaluating mean absolute errors (MAE) on the entire QM9 dataset. 
As shown in Fig.~\ref{fig:qm9}a, all properties exhibit a clean monotonic decrease in error with training set size, and the error curves are close to linear on a log-log scale.
The similarity of error convergence indicates that the model's learning behavior is largely property-agnostic in this regime.
HOMO and LUMO errors lie closely to each other across all training sizes, and the gap error follows the same trend at a slightly higher absolute value.
With 16000 training samples ($N=16000$), the model reaches MAEs of 21.2 meV for the HOMO energy, 21.2 meV for the LUMO energy, 28.4 meV for the gap, and 46.0 mDebye for the dipole moment.

Fig.~\ref{fig:qm9}b illustrates the learning curves for downfolding learning models across training sizes up to 32000.
Although the curves remain monotonically decreasing, the slopes are roughly half the values obtained from the direct learning model in cc-pVDZ basis. 
At the largest training set size evaluated ($N=32000$), MAEs of the downfolded Hamiltonian model (40.1 meV for the HOMO energy, 41.9 meV for the LUMO energy, 55.2 meV for the gap) remain noticeably larger, which relects the instrinc challenge in the downfolding learning due to the basis set mismatch between the input features and the ML targets.

Nevertheless, we note that the absolute accuracy of the downfolding learning model is competitive when compared against a natural physical reference. 
The HOMO-LUMO gap shifts by 157 meV in average when moving from the cc-pVDZ to the cc-pVTZ basis sets, indicating the typical magnitude of basis-set incompleteness error at the double $\zeta$ level. 
Across the full range of training sizes considered, the MAEs for all three energy targets lie well below this threshold (Fig. \ref{fig:qm9}b, dashed line). 
The residual error of the ML model trained on cc-pVTZ reference data is smaller than the systematic basis set error that would be incurred by using the cc-pVDZ basis in the underlying DFT calculations.

\subsubsection{Core-level carbon 1s energies}
Beyond frontier and valence observables, the accurate prediction of core-level binding energies is essential for applications in core-level spectroscopy \cite{golzeAccurateComputationalPrediction2022,gleasonCuXASNetRapidAccurate2025,foudaExperimentallyAccurateGraph2026}. 
We therefore evaluate the ability of the ML model to reproduce PBE0 C 1s eigenvalues on the QM9 dataset.
Notably, the training loss function did not explicitly penalize errors in the core orbital energies.

\begin{figure}[h]
    \centering
\includegraphics[width=0.7\linewidth]{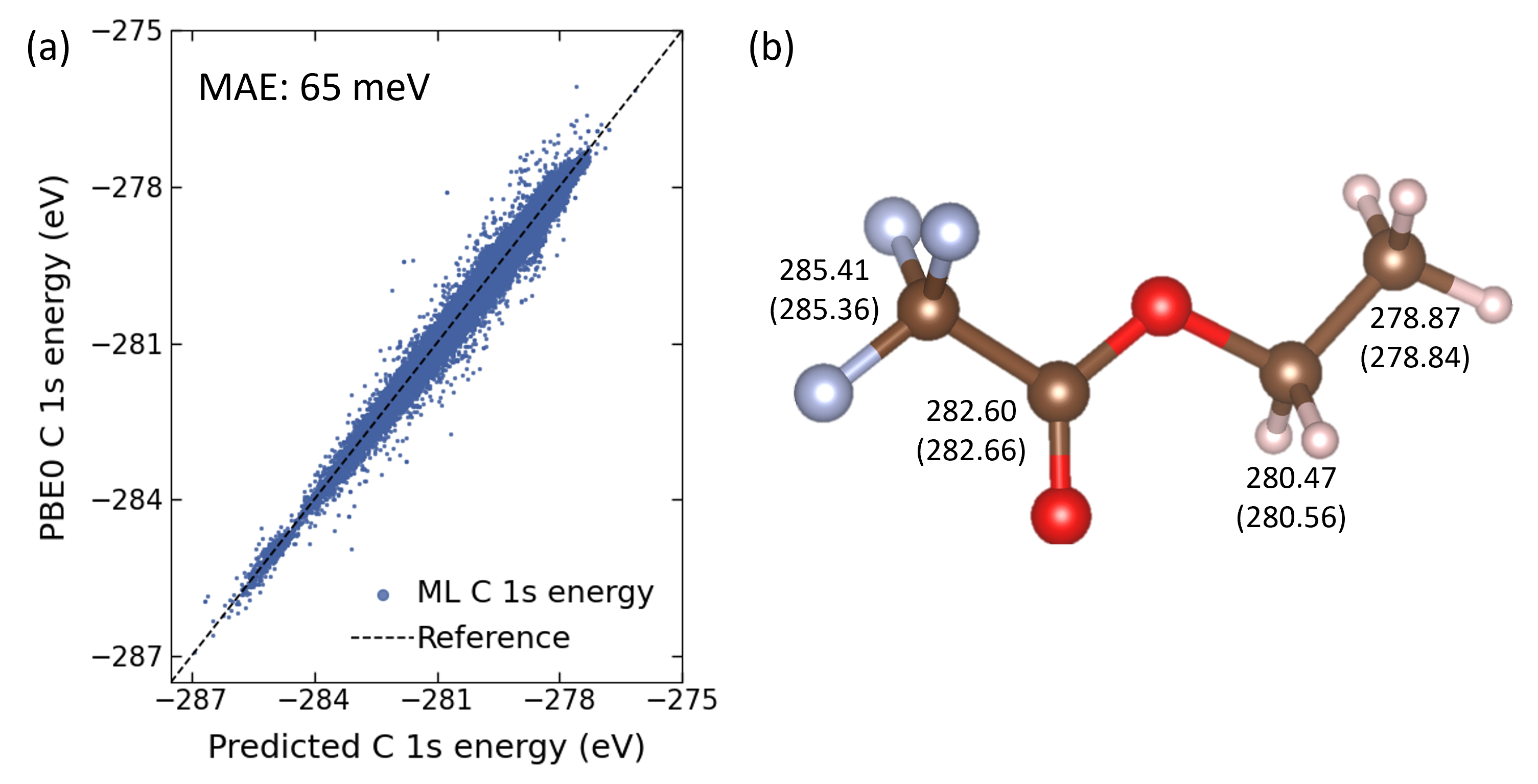}
\caption{
(a) ML-predicted carbon 1s core-level energies against PBE0 references for all carbon atoms in the QM9 test set. 
(b) ML-predicted carbon 1s binding energies for ethyl trifluoroacetate (the ESCA molecule) with reference PBE0 values shown in parentheses. All energies are given in eV.}
\label{fig:carbon1s}
\end{figure}

Results for the direct learning model ($N=16000$) are shown in Fig. \ref{fig:carbon1s}.
The predictions lie closely to the reference values across the full range spanned by the QM9 dataset, with a mean absolute error of 65 meV averaged over all carbon centers. 
The model therefore preserves chemical-environment descriptions that are sufficient for direct interpretation of spectroscopic signatures.
This is further illustrated by the ethyl trifluoroacetate (the ESCA molecule \cite{travnikovaESCAMoleculeHistorical2012}) example shown in Fig.~\ref{fig:carbon1s}b. 
The ESCA molecule contains four chemically distinct carbon sites whose C 1s binding energies span several electronvolts. The ML prediction reproduces all four binding energies within 0.1 eV of the reference values, demonstrating that the model captures both the absolute energy scale and the relative ordering of the core-level shifts.

\subsubsection{Density of states}
\begin{figure}[h]
    \centering
\includegraphics[width=1\linewidth]{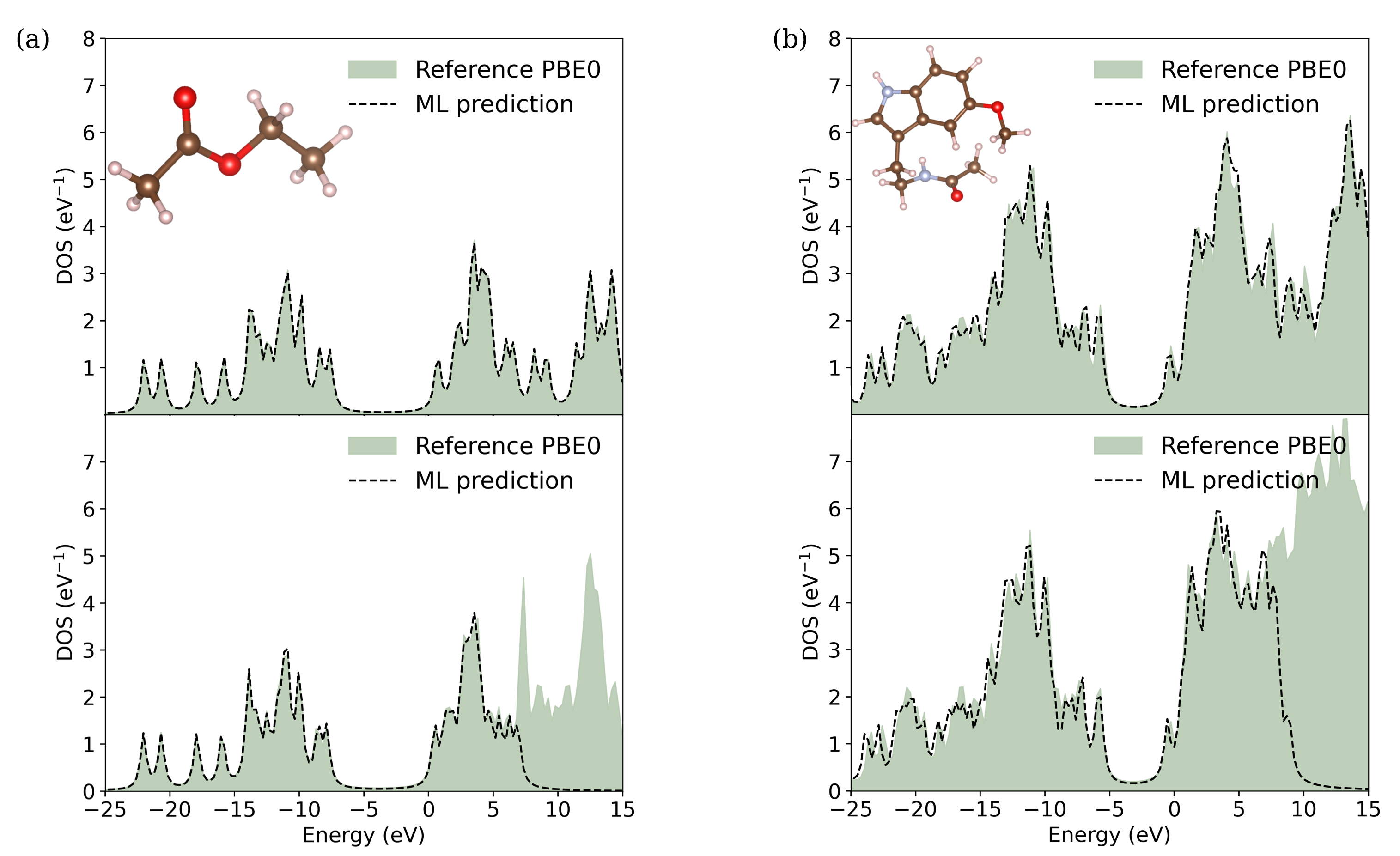}
\caption{Density of states (DOS) predicted by the ML model (dashed black) compared against the PBE0 reference (filled green). 
(a) ML prediction for ethyl acetate, a molecule contained within the QM9 dataset. (b) ML prediction for melatonin (\ch{C13H16N2O2}), a larger and chemically more complex molecule than QM9 molecules.
The top and bottom rows correspond to the direct learning and downfolding learning model results, respectively.}
\label{fig:dos}
\end{figure}

To assess prediction quality across the full molecular orbital energy spectrum, we compared the ML-predicted density of states (DOS) against the PBE0 reference for two representative molecules (Fig.~\ref{fig:dos}), using direct and downfolding learning models trained at $N=16000$ and $32000$, respectively. 
For ethyl acetate (Fig.~4a), both the direct and downfolding learning models reproduce the reference DOS with deviations that are not visually discernible.
The downfolding learning (bottom row), as expected, accurately reproduced the occupied and low-lying virtual orbital energies accessible within minimal basis sets.
For melatonin (Fig. \ref{fig:dos}b), which is roughly twice the size of typical QM9 molecules, the agreement remains robust.
Absolute ML errors of HOMO energy, LUMO energy, HOMO-LUMO gap are 0.10, 0.18, 0.08 eV for direct learning and 0.05, 0.16, 0.21 eV for downfolding model, respectively.
Taken together, these two case studies provide a complement to the results of frontier and core orbitals, showing both robustness and transferability of the ML models across the a wide energy window. 

\subsection{Intermolecular transfer integrals for organic semiconductors}
We trained direct learning models on the benzene dimer as a model system, alongside two representative charge-transport materials: tetracyanoquinodimethane (TCNQ) and tetrathiafulvalene (TTF).
For each system, we trained a model and the training data was generated by sampling molecular dynamics (MD) trajectories with constrained monomer-monomer distances and tilt angles.
A 2D-grid scan was performed for distances of $\{3.75,4.25,4.75,5.25,5.75\}$ {\AA} and tilt angles of $\{5,15,25,\dots,85\}$ degrees.
The training set comprised 1600 dimer geometries uniformly distributed across these distance-angle windows, supplemented by 400 monomer geometries extracted from free-monomer MD simulations. 
To properly capture intermolecular interactions, the CAM-B3LYP functional \cite{yanaiNewHybridExchange2004} was used to generate the reference DFT data. 
The training features and hyperparameters remained identical to those used for the QM9 dataset.

For the doubly degenerate HOMO orbitals in benzene, Eq.~\ref{eqn:tilde H} can be expanded to a $4\times4$ matrix, and the transfer integral is calculated as 
$t=\sqrt{\frac{1}{4}\sum_{I}\sum_{J}\left|t_{IJ}\right|^2}$, where $I,J$ denote the degenerate HOMO indices of the respective monomers.
The HOMO and LUMO were used to evaluate transfer integrals for TTF and TCNQ, respectively.
We evaluated model performance on fixed geometries at intermolecular distances of $\{4,5,6\}$ {\AA}, across a combination of tilt angles of $\{0,10,20,\dots,90\}$ degree and twisted angles.
Twisted angle spanned $\{10,20,30\}$ degree for benzene and $\{10,20,\dots,90\}$ degrees for TCNQ and TTF.
\begin{figure}[h]
    \centering
\includegraphics[width=1\linewidth]{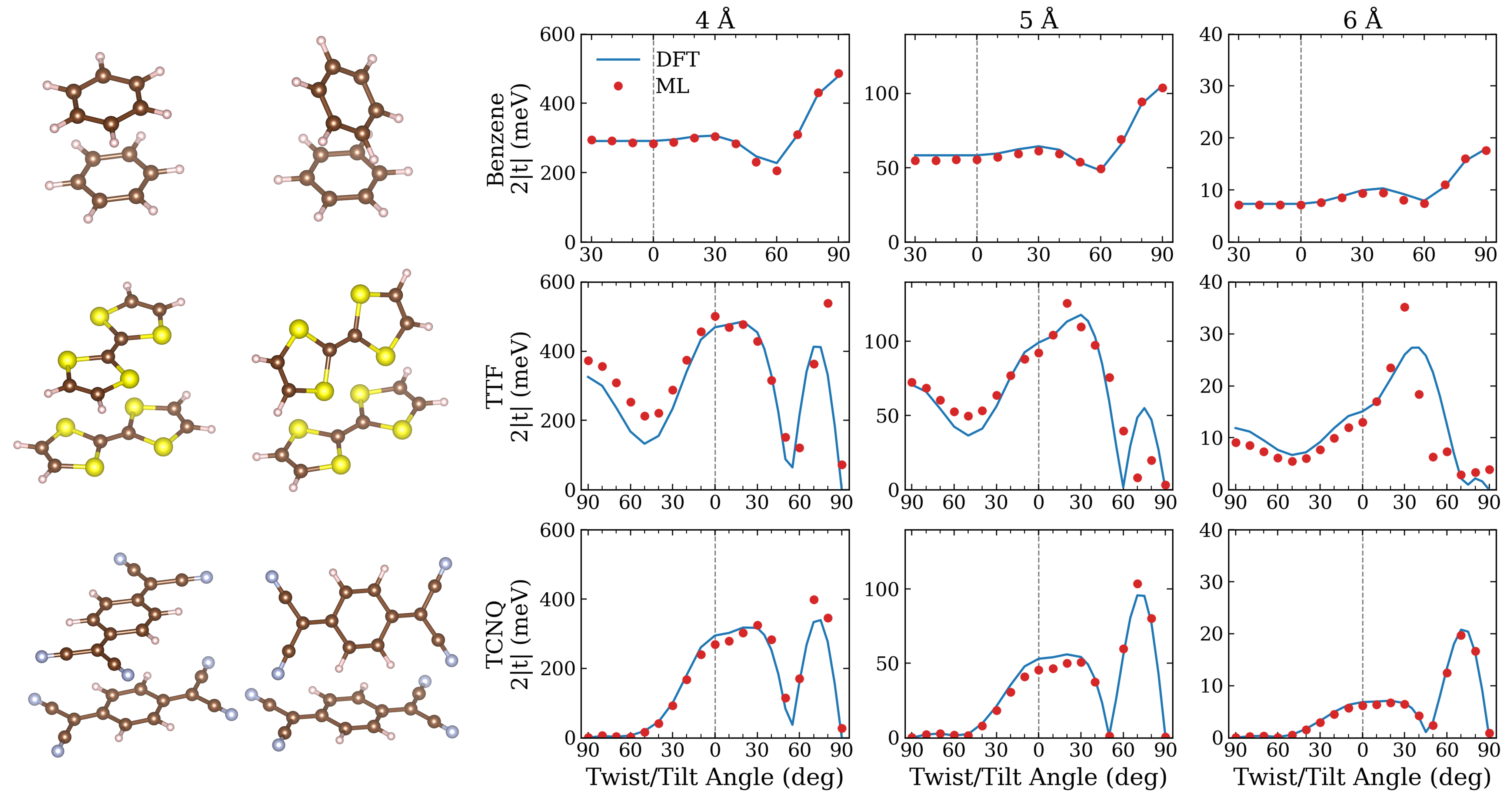}
\caption{Geometric dependence of the dimer transfer integral $2|t|$ for benzene (top), TTF (middle), and TCNQ (bottom), with representative twisted and tilted geometries shown on the left.
Columns on the right correspond to distances of 4, 5, and 6 {\AA}.
Each panel displays the QM reference (solid blue line) and the ML prediction (red dots) as a function of the twist or tilt angle. 
The dashed vertical line separates the twist (left) from the tilt (right) rotations.}
\label{fig:transfer_int}
\end{figure}

Test geometries were generated by manually stacking and rotating equilibrium monomer structures without further structural relaxation (Fig.~\ref{fig:transfer_int}).
The transfer integral $t$ depends sensitively on the frontier orbital overlaps and thus the relative orientation of two molecules.
This serves as a stringent test of whether the model has learned the true spatial structure of the underlying orbitals rather than merely their scalar energies.

The dimer transfer integral prediction results (reported as $2|t|$) are shown in Fig. \ref{fig:transfer_int}.
For all three molecular systems, the ML models quantitatively reproduce the decay of the transfer integrals with increasing intermolecular separation, consistent with the expected exponential decay of orbital overlap. The angular dependence of $|t|$ is also reproduced across all three distances. Quantitative agreement is best for benzene, with an overall MAE of 3.2 meV for $2|t|$, followed by TCNQ with an MAE of 7.5 meV. These errors are sufficiently small for realistic charge-transport modeling.
The TTF model slightly overestimates $|t|$ for all twisted geometries at 4 \AA. 
Specifically, a sharp peak at $80^\circ$ tilt angle (ML 538 meV vs.~DFT 332 meV)
causes a significantly larger MAE of 57.5 meV for 4 {\AA}, contributing to an overall MAE of 24.2 meV for TTF.
These larger deviations are tentatively attributed to the stronger intermolecular polarization of the sulfur valence orbitals in TTF dimers.

To demonstrate the transferability of the ML framework, we designed a substituted-benzene benchmark.
We expanded the training set by appending 10 isolated substituted benzene molecules: toluene (-CH3), fluorobenzene (-F), styrene (-CH=CH2), phenol (-OH), benzoic acid (-COOH), benzaldehyde (-CHO), aniline (-NH2), anisole (-OCH3), nitrobenzene (-NO2), benzonitrile (-CN).
Importantly, heterodimers consisting of benzene paired with a substituted benzene were entirely excluded from the training. 
These heterodimers introduce new intermolecular electrostatic environments and break the symmetry inherent to the homodimers.
Predicting transfer integrals for these heterodimers is thus an out-of-distribution task by construction.
\begin{figure}[h]
    \centering
\includegraphics[width=0.5\linewidth]{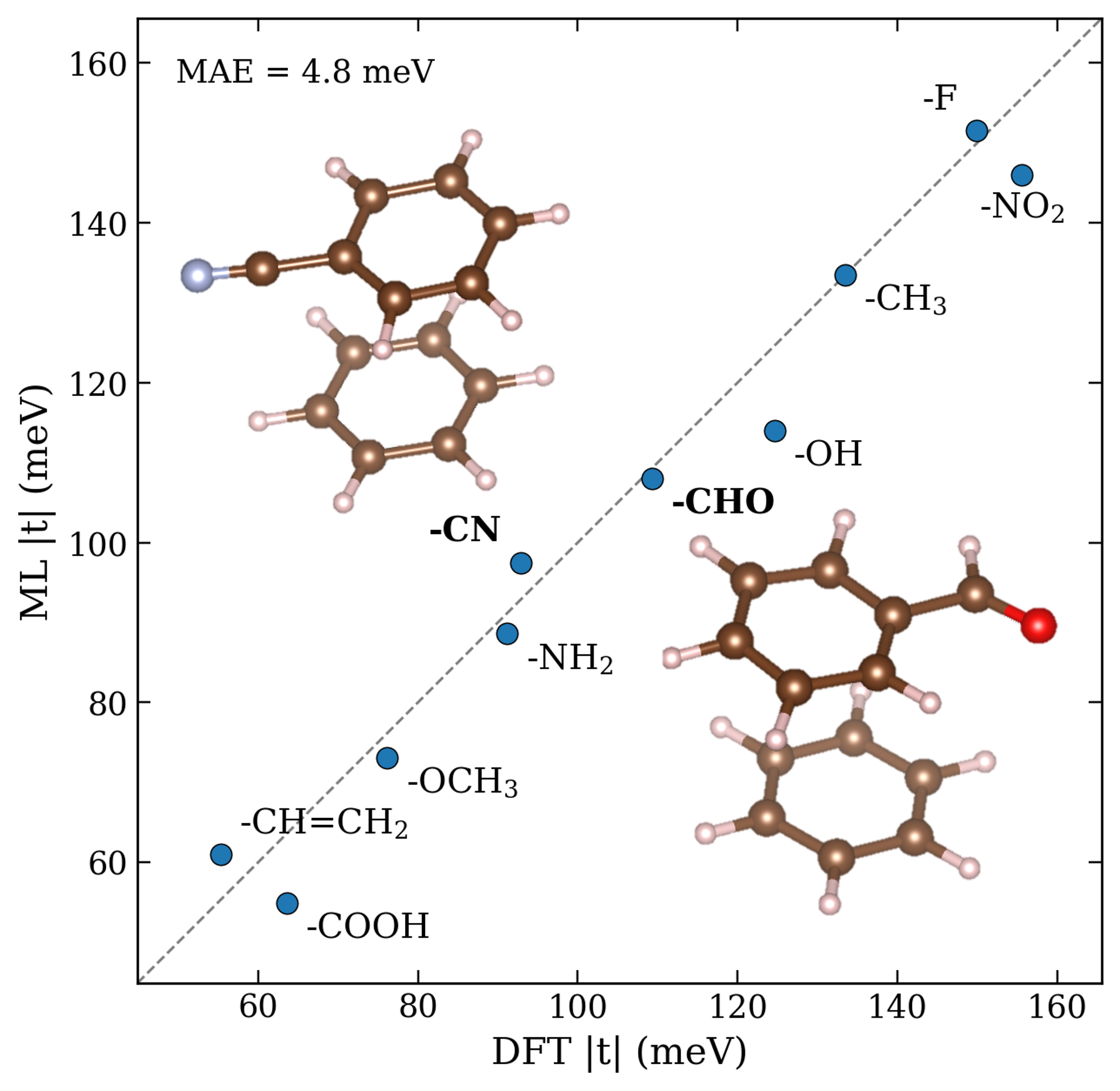}
\caption{Comparison of ML-predicted transfer integrals $|t|$ against reference DFT (PBE0) values.
Benzene-benzonitrile (-CN) and benzene-benzaldehyde (-CHO) dimer geometries are plotted along the corresponding data points.}
\label{fig:bz_deriv}
\end{figure}

The test set, shown in Fig.~\ref{fig:bz_deriv}, consists exclusively of the unseen heterodimers, evaluated at a stacked geometry with 4 {\AA} distance.
Despite the wide variance in coupling magnitude introduced by the different substituents, the ML predictions show excellent agreement with the DFT reference.
The MAE for $|t|$ is 4.8 meV, corresponding to a relative error of roughly 5\%, comparable to the in-distribution accuracy of the model on the homodimer angular benchmark.
Furthermore, the residuals exhibit no systematic bias as a function of substituent characters.
This confirms that the model has learned a highly compositional representation of intermolecular electronic coupling. 
The heterodimer coupling can be accurately assembled from the model's learned monomer orbitals without requiring the heterodimer itself to be present during training. 
From a practical standpoint, this transferability is valuable for high-throughput screening of organic semiconductor crystals, where the combinatorial growth of monomer-monomer pairings makes exhaustive data generation infeasible.

\section{Conclusion}
\label{summarySec}
We have introduced a transferable and scalable framework for learning DFT mean-field Hamiltonians from features derived from the SAP approximation. 
The SAP Hamiltonian provides an electronically informed, computationally efficient representation that captures electron-electron screening. 
These properties are utilized both to construct the symmetry-adapted learning basis and as input features for the graph neural network. 
A complementary downfolding scheme further enables the prediction of large-basis electronic structures directly from features with minimal-basis dimensions.

Evaluated on the QM9 dataset, our direct learning model reproduced PBE0 frontier-orbital energies, dipole moments, carbon 1s core-level shifts, and the density of states with high fidelity, while the downfolding model successfully recovered electronic spectra at cc-pVTZ quality. 
When applied to organic charge-transport materials, the models accurately predicted intermolecular transfer integrals for benzene, TTF, and TCNQ dimers across a broad spectrum of distances and orientations. 
Most notably, this accuracy successfully transferred to substituted-benzene heterodimers that were completely absent from the training data.

Future work will focus on extending this framework to periodic systems and to the downstream prediction of many-body Green's functions. 
When integrated into automated workflows, SAP-based Hamiltonian learning offers a  practical route toward accurate, high-throughput electronic structure prediction for molecular and materials discovery.

\begin{acknowledgement}
The work is supported by the National Science Foundation (Grant CHE-2337991). C.V. acknowledges support
from the Department of Defense through the National Defense
Science and Engineering Graduate (NDSEG) Fellowship Program. We thank the Yale Center for Research Computing for providing computing resources. 
\end{acknowledgement}
\clearpage
\bibliography{ref}
\clearpage
\end{document}